\documentclass[showpacs,amsmath,amssymb,nofootinbib,prd]{revtex4}

\usepackage{graphicx}
\usepackage{enumerate}

\newcommand{\beq}{\begin{equation}}
\newcommand{\eeq}{\end{equation}}
\newcommand{\bea}{\begin{eqnarray}}
\newcommand{\eea}{\end{eqnarray}}

\begin{document}

\title{Rare top decay $t \to c \bar{\textit{l}}\textit{l}$ as a probe of new physics}
\author{J.L. D\'{\i}az-Cruz}
\affiliation{Cuerpo Acad\'emico de
Part\'{\i}culas, Campos y Relatividad Facultad de Ciencias
F\'{\i}sico-Matem\'aticas, BUAP. Apdo. Postal 1364, C.P. 72000
Puebla, Pue., M\'exico}
\author{A. Diaz-Furlong}
\affiliation{Cuerpo Acad\'emico de
Part\'{\i}culas, Campos y Relatividad Facultad de Ciencias
F\'{\i}sico-Matem\'aticas, BUAP. Apdo. Postal 1364, C.P. 72000
Puebla, Pue., M\'exico}
\author{R. Gait\'an-Lozano}
\affiliation{Departamento de F\'isica, FES-Cuautitlan,
UNAM, C.P. 54770, Estado de M\'exico, M\'exico}
\author{J.H. Montes de Oca Y.}
\email{josehalim@comunidad.unam.mx} \affiliation{Departamento de
F\'isica, FES-Cuautitlan, UNAM, C.P. 54770, Estado de M\'exico,
M\'exico}

\begin{abstract}
The rare top decay $t \to c \bar{\textit{l}}\textit{l}$, which
involves flavor violation, is studied as a possible probe of new
physics. This decay is analyzed with the simplest Standard Model
extensions with additional gauge symmetry formalism. The considered
extension is the Left-Right Symmetric Model, including a new neutral
gauge boson $Z^\prime$ that allows to obtain the decay at tree level
through Flavor Changing Neutral Currents (FCNC) couplings. The
neutral gauge boson couplings are considered diagonal but family
non-universal in order to induce these FCNC. We find the $BR(t \to c
\bar{\textit{l}}\textit{l})\sim10^{-13}$ for a range $1$ TeV$ \leq
M_{Z^\prime} \leq$ $3$ TeV.

\end{abstract}


\maketitle

\section{Introduction}

The last year we have witnessed the impressive work of LHC, which
has reached a luminosity that has allowed to test the Standard Model
(SM) at extraordinary levels \cite{luminosidad}. In particular, LHC
has provided notable bounds on the SM Higgs boson mass \cite{LHCmass}. After the
discovery of the top quark at Fermilab Tevatron Collider,
experimental attention has been turned on the examination of its
production mechanisms and decay properties. Within the SM, the top
quark production cross section is evaluated with an uncertainty of
the order of $\sim 15\%$, while it is assumed to decay to a $W$
boson and a $b$ quark almost $100\%$ of the time. With higher
energy, as planned, the LHC will also become an amazing top factory,
allowing to test the top properties, its couplings to SM channels and
rare decays. Because about $10^7-10^8$ top pairs will be produced
per year, rare decays with B.R. of order $10^{-5}-10^{-6}$ may be
detectable, depending on the signal. For the $W$ boson coupling to
fermion pairs ($td_iW^\pm$), the structure is proportional to the
CKM element $V_{td_i}$ in the framework of the SM. Therefore, the
decay $t\to b+W$ dominates its branching ratio. Radiative
corrections to this mode are of order $10\%$ and are difficult to
detect at hadron collider, but may be at the reach of the International Lineal Collider (ILC). Top
quark decays from Flavor Changing Neutral Currents (FCNC), such as
$t\to c \gamma$, $t\to c g$, $t\to c Z$ and $t\to c \phi$, have been
studied, both in the context of the SM and new physics. In the SM,
the branching ratio of FCNC top decays is extremely suppressed. The
rare top quark decay $t\to c+\gamma$ was calculated \cite{topfoton},
the result implied a suppressed branching ratio, less than about
$10^{-10}$, which was confirmed when subsequent analysis, included
the correct top mass value, and gave $BR(t \to c+\gamma)= 5 \times
10^{-13}$ \cite{Atwood:2005bf}. The decays $t\to c+Z$ and $t\to c+g$
were also calculated \cite{topcharm},\cite{Abazov:2011qf}. The resulting branching ratios
obtained there turned out to be $BR(t\to c+Z)=1.3 \times 10^{-13}$
and $BR(t\to c+g)=5 \times 10^{-11}$. The top-charm coupling with
the SM Higgs  $\phi^0$ could be induced at one-loop level with a
resulting branching ratio $BR(t\to c+\phi^0)=10^{-15}$. The FCNC top
decays involving a pair of vector bosons in the final state, $t \to
cVV$, can also be of interest \cite{vectoriales}. Although one could
expect such modes to be even more suppressed than the ones with a
single vector boson, the appearance of an intermediate scalar
resonance, as in the previous case, could enhance the branching
ratio. Furthermore, it also seems possible to allow the tree-level
decay  $t\to b+WZ$, at least close to threshold, because of the
large top quark mass \cite{dennys}. The top decay into the light quarks $t\to
W+d(s)$ is suppressed, as they are proportional to $V_{td(s)}$
\cite{lorenzo-gaitan}. Probably for this reason, the SM corrections
to this mode have not been studied, though the QCD corrections
should be the similar for both modes. However, it may be possible to
get a large enhancement that could even make it detectable at the
ILC in extensions of the SM. Some typical results for the top decays
in the SM are summarized in Table I. This table also includes, for
comparison, the results for top branching ratios from models beyond
the SM, in particular from the THDM-III \cite{diaz1} and SUSY
\cite{diaz2}, which will be discussed in this work. Another
interesting mode is the decay $t\rightarrow c l^+l^-$, which could
be mediated by a vector resonance. Within the SM one could expect a
$BR(t\rightarrow cl^+l^-)\approx BR(t\rightarrow cZ)\cdot
BR(Z\rightarrow l^+l^-)<10^{-10}$. Thus, this mode offers the
possibility to test extensions of the SM that include an additional
vector boson $Z^\prime$, which could have SM-like couplings to
$l^+l^-$, but enhanced coupling $tcZ^\prime$. In this paper, we
evaluate this decay mode within a particular extension that are well
motivated and produce interesting signals. The so-called Left-Right
Symmetric Model with Non-Universal extra gauge bosons, where there
are strong constraints for transitions involving the 1st and 2nd
generations, but admit larger effects for transitions involving the
3rd generation. We find that one can obtain $BR(t\rightarrow
cl^+l^-)<10^{-10}$. This paper is organized as follows. In section
2, the parametrization of the couplings of  $Z^\prime$ neutral gauge
boson and evaluation the decay width for $t\rightarrow c l^+l^-$. In
section 3, the relevant details of the Non-universal Z model and
evaluation of the corresponding $BR(t\rightarrow cl^+l^-)$. Finally,
our conclusions appear in section 4.

\begin{table}
\begin{center}
\begin{tabular}{| c| c| c| c| }
\hline\hline
{\bf BR\ } & SM & THDM-III & MSSM \\
\hline $\mathbf{BR(t \to sW)}$ & $2.2 \times 10^{-3}$ & $\sim
10^{-3}$ &
$10^{-3}-10^{-2}$ \\
\hline $\mathbf{BR(t \to c\phi^0)}$ & $10^{-13}-10^{-15}$ & $\sim
10^{-2}$ &
$10^{-5}-10^{-4}$\\
\hline
$\mathbf{BR(t \to c \gamma)}$ & $5\times 10^{-13}$ & $< 10^{-6}$ &  $<10^{-7}$ \\
\hline
$\mathbf{BR(t \to c Z)}$ & $1.3 \times 10^{-13}$ &  $< 10^{-6}$ & $<10^{-7}$ \\
\hline
$\mathbf{BR(t \to c g)}$ & $5\times 10^{-11}$ & $< 10^{-6}$ & $< 10^{-5}$ \\
\hline
$\mathbf{BR(t \to c \gamma\gamma)}$ & $<10^{-16}$ & $\sim 10^{-4}$ &  $<10^{-8}$ \\
\hline $\mathbf{BR(t \to c WW)}$ & $2\times 10^{-13}$  &
$10^{-4}-10^{-3}$ &
??  \\
\hline
$\mathbf{BR(t \to cZZ)}$ & -- & $10^{-5}-10^{-3}$ & ?? \\
\hline
$\mathbf{BR(t \to bWZ)}$ & $2\times 10^{-6}$ & $\simeq 10^{-4}$ & ?? \\
\hline
\end{tabular}
\end{center}
\vspace{-0.55cm} \caption{ Branching ratios for some CKM-suppressed
and FCNC top quark decays in the SM and beyond, for $m_t=173.5-178$
GeV. Decays into a pair of massive gauge bosons include  finite
width effects of final state unstable particles \cite{lorenzo-gaitan}, \cite{diaz1}.}
\vspace{-0.4cm}
\end{table}

\section{Flavor-changing neutral currents from family non-universal couplings}

The extension for Standard Model (SM) known as Left-Right symmetric
model (LRSM) is considered in order to include extra gauge bosons.
The gauge symmetry group of the LRSM is $SU(2)_L \otimes
SU(2)_{R}\otimes U(1)_{B-L}$ \cite{Mohapatra-LRSM}. In literature,
several models contain extra neutral gauge bosons through increasing
the gauge symmetry group \cite{Langacker-Zprime} but LRSM is the
simplest model based on physical motivation. In this work, exotic
fermions are not included in fermion field contain, only the SM
fermions.

The notation for parameters and formalism introduced by Langacker and
Plumacher have been used in this work \cite{Langacker-plumer}. Then,
the couplings for neutral gauge bosons with fermions are given by
\begin{equation}
-\mathcal{L}_{NC}=eJ_{EM}^{\mu }A_{\mu
}+\sum_{a=1}^{2}g_{a}J_{a}^{\mu }Z_{a\mu },
\end{equation}
where $Z_{1\mu }$ is the usual electroweak neutral gauge boson,
$Z_{2\mu }$ is the neutral gauge boson associated with the
additional gauge symmetry and $g_{1,2}$ are their respective gauge
coupling constants. The $Z_{a\mu}$ and $J_{a\mu}$ are written in
gauge eigenstate basis. The general form of the $J_{2\mu}$ current
is
\begin{equation}
J_{2\mu }=\sum_{h}\sum_{i,j}\overline{f}_{i}^{0}\gamma _{\mu
}\epsilon _{hij}^f P_{h}f_{j}^{0}, \label{j2}
\end{equation}
where $i,j= 1,2,3$, $h=L,R$ and $P_{L,R}=(1\mp\gamma_5)/2$.
$\epsilon _ {hij}^f$ are model depending parameters described below.
FCNCs can be introduced through the $\epsilon _ {hij}^f$ parameters,
when family non-universal assumption is done. The flavor diagonal
and family universal couplings mean that $\epsilon _
{hij}^f=Q_h^f\delta_{ij}$, where $Q_h^f$ denotes the chiral charges
and depends on the model. For LRSM, the chiral charges are given by
\begin{equation}
Q_L^i=-\sqrt{\frac{3}{5}}\left( \frac{1}{2\alpha}\right)(B-L)_i
\label{QiL}
\end{equation}
and
\begin{equation}
Q_R^i=\sqrt{\frac{3}{5}}\left( \alpha
T_{3R}^i-\frac{1}{2\alpha}(B-L)_i \right), \label{QiR}
\end{equation}
where $B$ and $L$ denote the baryon and lepton numbers of the
fermion $i$, respectively. $T_{3R}$ is the third component of its
right-handed isospin in the $SU(2)_R$ group and
$\alpha^2=(1-2\sin^2\theta_W)/\sin^2\theta_W$. If the $Z_2$
couplings are diagonal but they are family non-universal, then
$\epsilon _ {hij}^f= x ^f_{hi} Q_h^q\delta_{ij}$,here the repeated
indexes do not denoted a sum. After changing to mass eigenstate
basis, the up-quark sector current is
\begin{equation}
J^u_{2\mu}=\sum_h \left( \bar{u},\, \bar{c},\,
\bar{t}\right)\gamma_\mu V^\dagger_{uh} \epsilon_{h}^u V_{uh} P_h
\left(
\begin{array}{c}
u\\
c\\
t\\
\end{array}
\right).
\end{equation}
Analogously in the case of the down-quark sector. However, for
simplicity, we shall assume that down-quark sector has
non-mixing, then left-handed and right-handed CKM matrix can be
$V_{CKM}\approx V_{uL}$ and $V_R\approx V_{uR}$, respectively.
Therefore, the following $3\times3$ matrices are defined as
\begin{equation}
B^u_{L}=V^\dagger_{CKM}\epsilon^u_L V_{CKM}
\end{equation}
and
\begin{equation}
B^u_{R}=V^\dagger_{R}\epsilon^u_R V_{R}.
\end{equation}
The usual and known CKM matrix parametrization in terms of the
Wolfenstein parameters, $A$, $\lambda$, $\eta$, $\delta$ are used in
this work \cite{Wolfenstein1983}. For $V_R$ matrix, the
parametrization given by Zhang et. al. is taken into account
\cite{Mohapatra-CKM-right},
\begin{equation}
V_R=P_U \widetilde{V}_L P_D,
\end{equation}
in which
$P_U=\textrm{diag}\left(s_u,\,s_ce^{2i\theta_2},\,s_te^{2i\theta_3}
\right)$,
$P_D=\textrm{diag}\left(s_de^{i\theta_1},\,s_se^{-i\theta_2},\,s_be^{-i\theta_3}
\right)$ and
\begin{equation}
\widetilde{V}_L =\left(
                   \begin{array}{ccc}
                     1-\frac{\lambda^2}{2} & \lambda & A\lambda^3(\rho-i\eta) \\
                     -\lambda & 1-\frac{\lambda^2}{2} & A\lambda^2e^{-2i\theta_2} \\
                     A\lambda^3(1-\rho-i\eta)& -A\lambda^2e^{2i\theta_2} & 1 \\
                   \end{array}
                 \right).
\end{equation}
The $s_q$ guarantee that up-type and down-type quark mass matrix
elements are positive. The $\theta_i$ phases come up as well as it
happens in CKM matrix.

Finally, electroweak and extra neutral gauge bosons are mixed by
$\xi$ parameter. Since $\xi<<1$, $Z_1$ and $Z_2$ can be identified
with physical gauge bosons $Z$ and $Z^\prime$, respectively.

\section{Rare top decay $t \to c \bar{\textit{l}}\textit{l}$}

\subsection{Decay width for $t \to c \bar{\textit{l}}\textit{l}$}

\begin{figure}[tbp]
\centering
\includegraphics[scale=0.75]{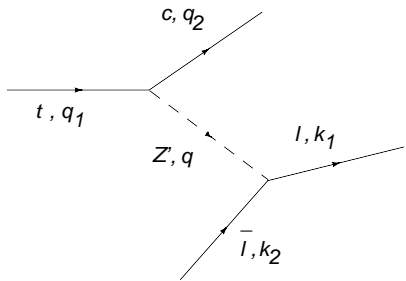}
\caption{Tree level Feynman diagram for rare top decay mediated by
$Z^\prime$ extra gauge boson.} \label{d1}
\end{figure}
The branching ratio for the rare top decay $t \to c
\bar{\textit{l}}\textit{l}$ is calculated by using the formalism
introduced in the previous section. The diagram for this decay is
shown in figure \ref{d1}. If $q^{2}<<M_{Z^{\prime }}^{2}$, the
average of amplitude as
\begin{equation}
\left\vert \overline{\mathcal{M}}_{t\rightarrow c\overline{l}l}
\right\vert ^{2}=\frac{g_1^4}{ 8M_{Z^{\prime }}^{4}}\left[
B_{1}\left( q_{1}\cdot k_{1}\right) \left( q_{2}\cdot k_{2}\right)
+B_{2}\left( q_{1}\cdot k_{2}\right) \left( q_{2}\cdot k_{1}\right)
+B_{3}m_{l}^{2}\left( q_{1}\cdot q_{2}\right)
-B_{4}m_{t}m_{c}\left( k_{1}\cdot k_{2}\right) -B_{5}m_{t}m_{c}m_{l}^{2}%
\right] ,
\end{equation}
and then the decay width becomes
\begin{equation}
\Gamma _{t\longrightarrow c\overline{l}l}
=\frac{g^{4}m_{t}^{5}}{\left( 16\pi \right) ^{3}M_{Z^{\prime
}}^{4}}\sum_{i=1}^5a_iB_i,
\end{equation}
where the $a_i$ and $B_i$ are explicitly written in the appendix. 

\subsection{Parameter space}

Three scenarios are analyzed in order to
explore the behavior of the family non-universal parameters. All
scenarios assume that the $Z^\prime$ coupling to the leptons and
down-type quarks are flavor diagonal and family universal, that is,
$\left(x^{d}_{L,R}\right)_{ij}=\left(x^{l}_{L,R}\right)_{ij}=\delta_{ij}$.
The scenarios are:
\begin{enumerate}[i)]
  \item \emph{Left-handed family non-universality for up-type quarks}. The right-handed up-type quarks are family
  universal and only the last family for
  left up-type quarks has family non-universal coupling with
  $Z^\prime$ \cite{Arhrib},
  $\left(x^{u}_{R}\right)_{ij}=\delta_{ij}$,
  $\left(x^{u}_{L}\right)_{11}=\left(x^{u}_{L}\right)_{22}=1$ and
  $\left(x^{u}_{L}\right)_{33}=x$. The parameter $x$ must be close
  to 1, but it is not exactly 1 in order to obtain FCNC from $Z^\prime$ boson.
  \item \emph{Left-Right handed family non-universality for up-type
  quarks}. The third generation of the right-handed and left-handed up-type quarks are family non-universal,
  $\left(x^{u}_{L,R}\right)_{11}=\left(x^{u}_{L,R}\right)_{22}=1$ and
  $\left(x^{u}_{L,R}\right)_{33}=x$.
  \item \emph{General left-handed family non-universality for up-type
  quarks}. Three families of the left-handed up-type quarks have family non-universal couplings
  $\left(x^{u}_{L}\right)_{11}\equiv x_1$, $\left(x^{u}_{L}\right)_{22}\equiv x_2$ and $\left(x^{u}_{L}\right)_{33}\equiv x_3$.
  As in first scenario, right-handed up-type quarks are family universal.
\end{enumerate}

\subsection{Numerical result and discussion}
\begin{figure}[tbp]
\centering
\includegraphics[scale=0.6]{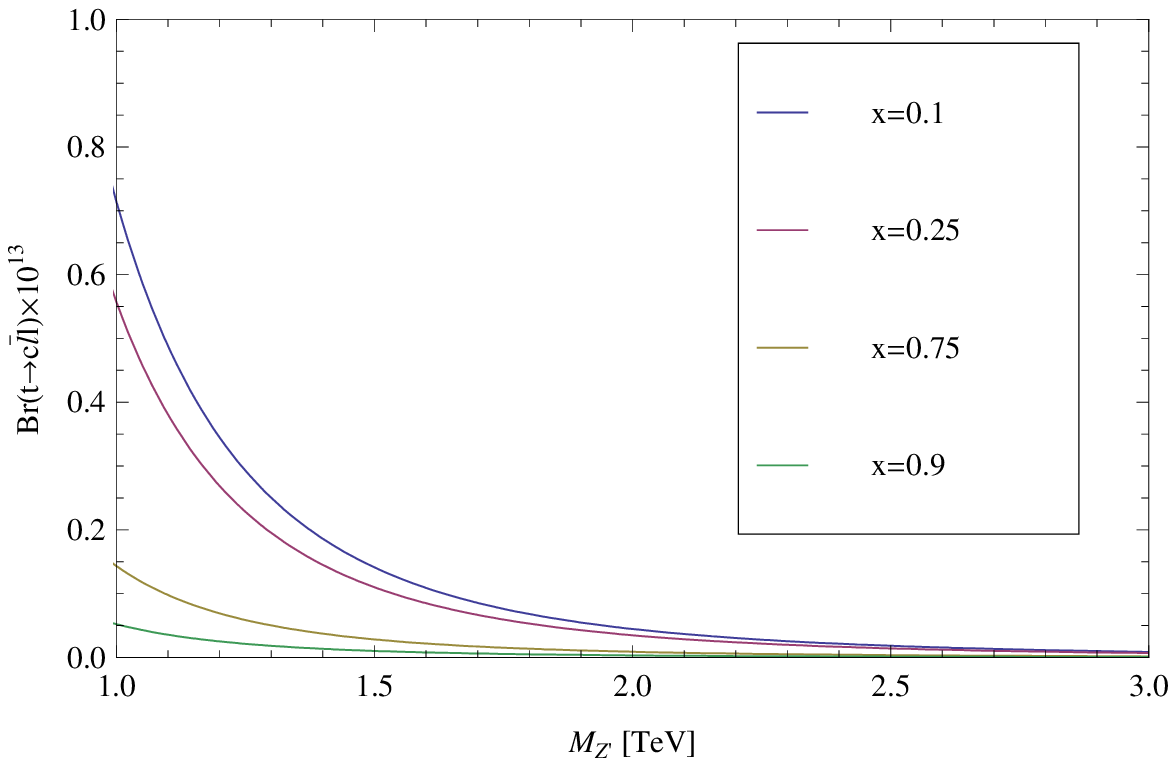}
\hspace{1cm}
\includegraphics[scale=0.6]{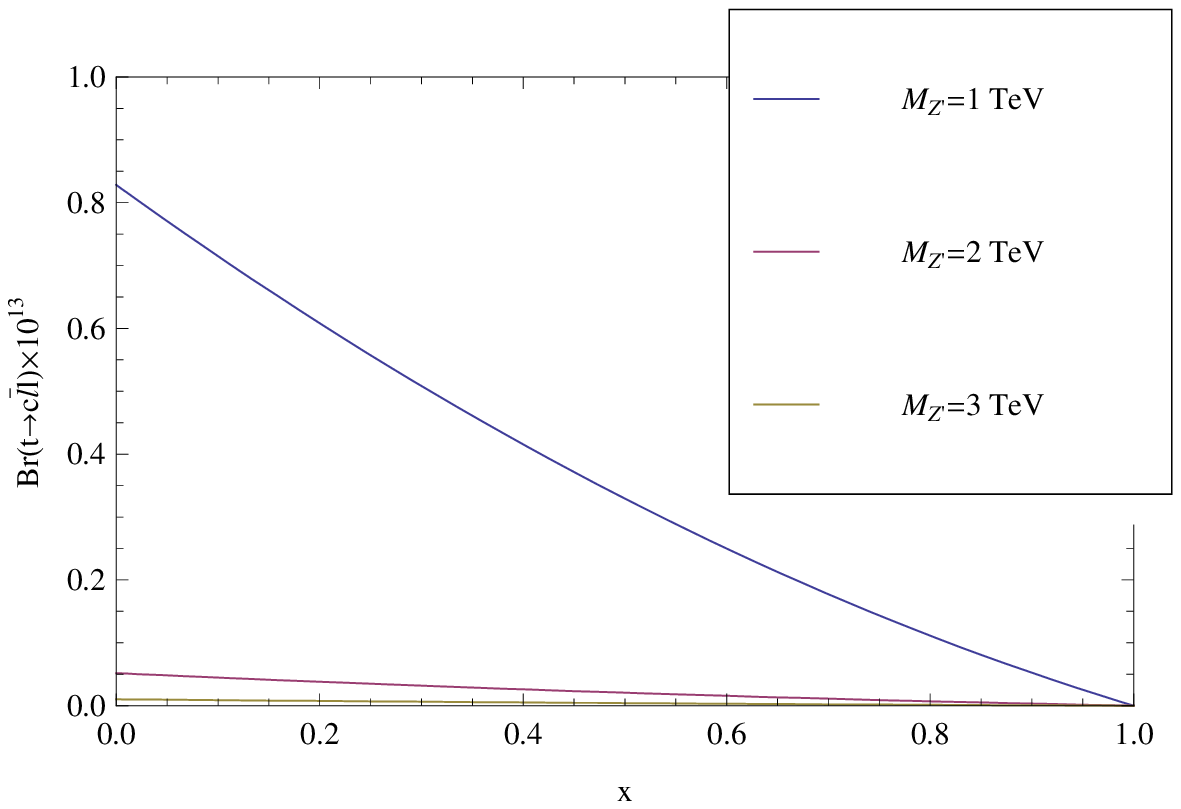}
\caption{Left(right) figure corresponds to branching ratio as
function of the $x$ ($Z^\prime$ mass) for the scenario i).}
\label{f1}
\end{figure}
\begin{figure}[tbp]
\centering
\includegraphics[scale=0.6]{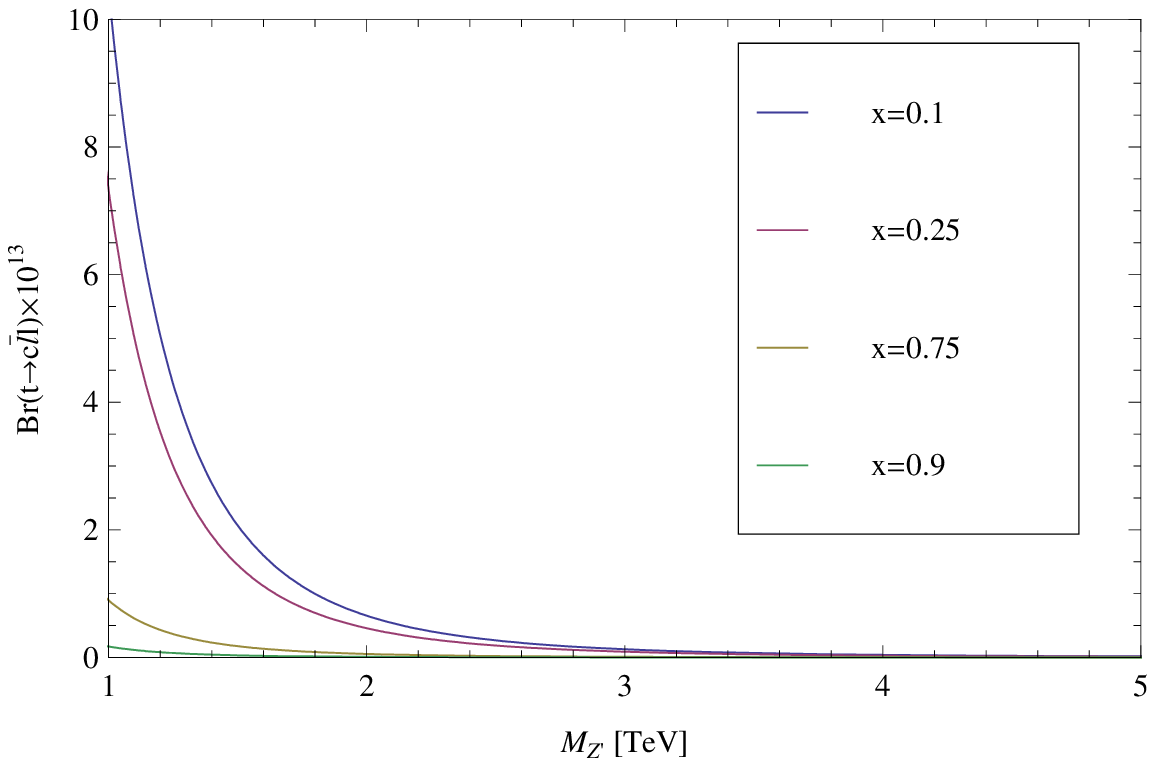}
\hspace{1cm}
\includegraphics[scale=0.6]{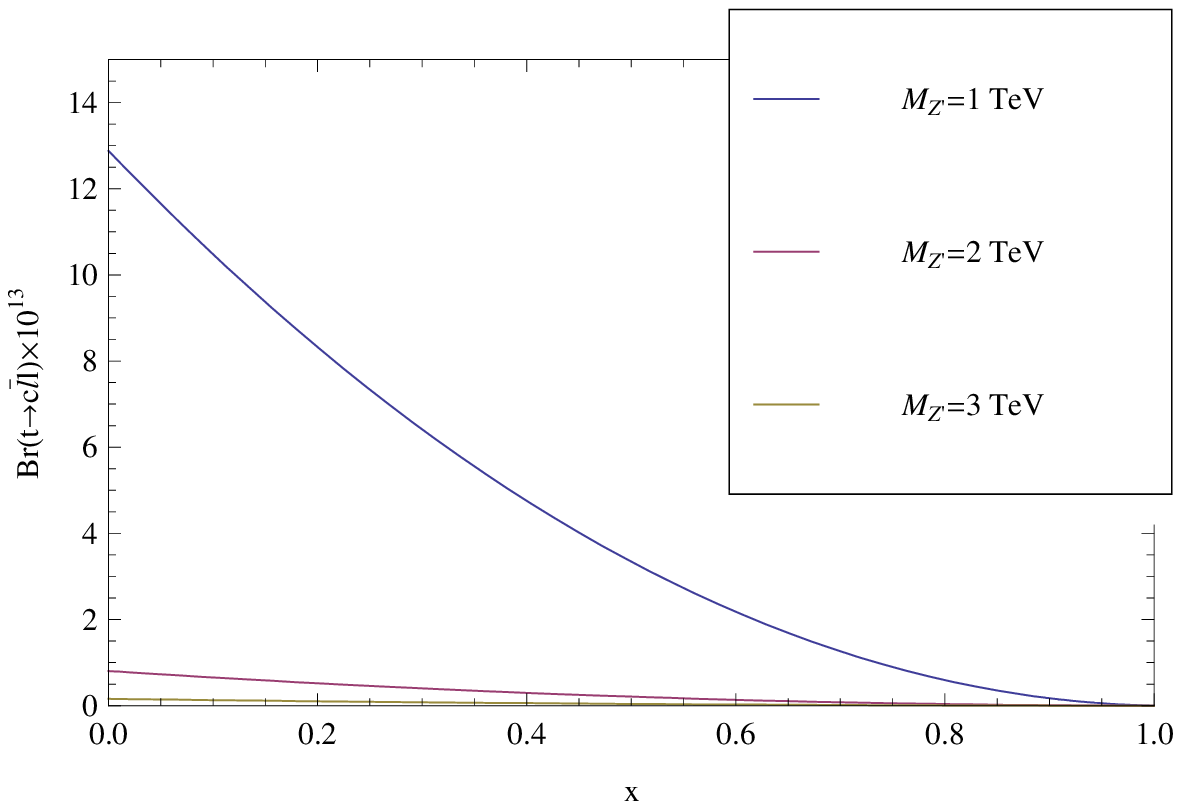}
\caption{Left(right) figure corresponds to branching ratio as
function of the $x$ ($Z^\prime$ mass) for the scenario ii).}
\label{f2}
\end{figure}
\begin{figure}[tbp]
\centering
\includegraphics[scale=0.7]{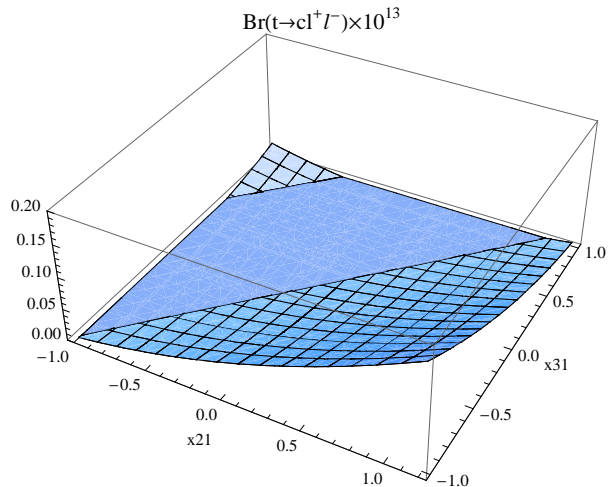}
\caption{Branching ratio for the scenario iii).} \label{f3}
\end{figure}
\begin{figure}[tbp]
\centering
\includegraphics[scale=0.7]{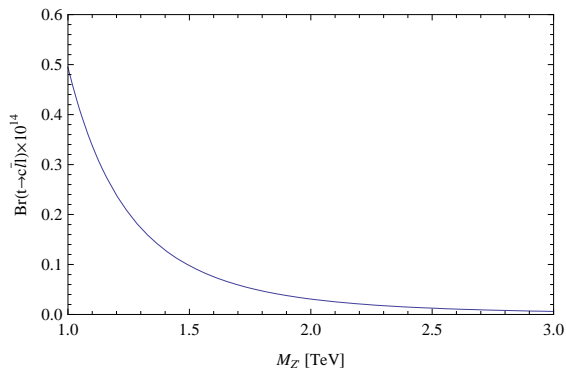}
\caption{Example branching ratio in the scenario iii) when allowed
values $x_1=0.8$, $x_1=1$, $x_3=0.9$ are considered.} \label{ex}
\end{figure}
We have taken from Particle Data Group the central values of the SM
parameters \cite{pdg}. Then, numerical values of the chiral charges
are shown in table \ref{tableQ}. For all considered scenarios in
previous section, appendix \ref{appendix1} contains the analytical
expressions and numerical values of the $B^{u,l}_{L,R}$ parameters
and $I_i$ integrals.

In the case of the scenario i) and ii), the branching ratio of the
rare top decay $t\rightarrow c \bar{l}\, l$ has been obtained as
function of the family non-universal parameter $x$ and the
$Z^\prime$ mass, see figures \ref{f1} and \ref{f2}. In the case of
the scenario iii), the branching ratio depends on the parameters
$x_{31}=x_3-x_1$ and $x_{32}=x_3-x_2$. The figure \ref{f3} is
obtained  by using the representative lower bound of the $Z^\prime$
mass, which is around $1$ TeV \cite{pdg}.
\begin{table}
\centering
\begin{tabular}{|c|c|}
  \hline
  Chiral charge & LRSM \\
  \hline
  $Q_L^u$ &-0.0847 \\
  \hline
  $Q_R^u$ &0.5048 \\
  \hline
  $Q_L^d$ &-0.0847 \\
  \hline
  $Q_R^d$ &-0.6744 \\
  \hline
  $Q_L^e$ &0.2543 \\
  \hline
  $Q_R^e$ &-0.3352 \\
  \hline
\end{tabular}
  \caption{Numerical values of the chiral charges for leptons and quarks obtained by using (\ref{QiL}), (\ref{QiR}) and $\sin^2\theta_W=0.2316$.}
  \label{tableQ}
\end{table}
As mentioned before, the family non-universal couplings are
important to including FCNC from extra gauge bosons. In both
scenarios i) and ii), branching ratio and family non-universal
parameter shows a decreasing behavior for values of the $Z^\prime$
mass greater than 1 TeV. However, the results for scenario ii) are
10 times bigger than the results for scenario i). A prominent
feature of right-handed CKM matrix obtained in the branching ratio
is its the increment.

We finally discuss the branching ratio in the scenario iii). The
branching ratio not only depends on two parameters as above, but two
more parameters are added for the general family non-universality.
However, the unitary of the CKM matrix allows us to write the
branching ratio as function of $x_{3,1}=x_3-x_1$ and
$x_{2,1}=x_2-x_1$. The domain of the $\left(x_{21},\,x_{31}\right)$
shall be $\left[0,1\right]\times\left[0,1\right]$ in order to have
closed values between the family non-universal parameters,
$x_{1,2,3}$, and near to 1. The not allowed region in the domain is
bounded among line $x_{31}=0.9746x_{21}$ and
$x_{31}=0.9746x_{21}+1.4404$. Out of this region, there still exit
some values of the $x_{31}$ and $x_{21}$ for which the $x_{1,2,3}$
could be negative. It can be control when we constrain the
parameters as $x_3>x_{31}$ and $x_{21}>x_{31}-x_3$. Figure \ref{ex}
shows an example when $x_{31}=0.1$ and $x_{21}=0.2$ are selected.
Finally, we can remark that the branching ratio of the rare top
decay is very suppressed in all scenario.
\section{Conclusion}
We find the branching ratio of the rare top decay at tree level for
three posible scenarios. The branching ratio is between
$10\times^{-13}$ $\sim 10\times^{-12}$. The right-handed CKM matrix
contribution in scenario ii) helps to give an arise in the branching
ration. However, for any scenario is still very suppressed.

We can also obtain a allowed region for the family non-universal
parameters in scenario iii). The allowed parameters keep positive
the branching ratio.

\noindent\textbf{Acknowledgments.}

\noindent
This work is supported in part by PAPIIT project IN117611-3, Sistema
Nacional de Investigadores (SNI) in M\'exico. J.H.Montes de Oca Y.
is thankful for support from the postdoctoral DGAPA-UNAM grant.

\appendix

\section{Analytical expressions and numerical values}
\label{appendix1}

Below we give the complete analytic formulae for the family
non-universal parameters in three considered scenarios. We also
present the analytic expression of the integrals $I_{1,...,5}$.

\subsection{Family non-universal couplings}

For the scenario i)
\begin{equation}
\left[ B_{L}^{u}\right] _{32}=Q_{L}^{u}\left( x-1\right)
V_{tb}V_{cb}^{\ast },
\end{equation}
\begin{equation}
\left[ B_{R}^{u}\right] _{32}=0,
\end{equation}
\begin{equation}
\left[ B_{L,R}^{l}\right] _{33}=Q_{L,R}^{l}.
\end{equation}
Then, we write last the expressions of the $B_{1,...,5}$ in terms of
the Wolfeinstein parameter, $V_{tb}=1$ $\ $and $V_{cb}=A\lambda
^{2}$,
\begin{equation}
B_{1}=2\left( Q_{L}^{u}Q_{L}^{l}\right) ^{2}\left( x-1\right)
^{2}A^{2}\lambda ^{4}
\end{equation}
\begin{equation}
B_{2}=2\left( Q_{L}^{u}Q_{R}^{l}\right) ^{2}\left( x-1\right)
^{2}A^{2}\lambda ^{4}
\end{equation}
\begin{equation}
 B_{3}=0
\end{equation}
\begin{equation}
B_{4}=Q_{R}^{l}Q_{L}^{u}\left( x-1\right) A\lambda ^{2}\left(
\left\vert Q_{L}^{l}\right\vert ^{2}+\left\vert Q_{R}^{l}\right\vert
^{2}\right)
\end{equation}
and
\begin{equation}
B_{5}=0.
\end{equation}
For the scenario ii)
\begin{equation}
\left[ B_{L}^{u}\right] _{32}=Q_{L}^{u}\left( x-1\right) A\lambda
^{2},
\end{equation}
\begin{equation}
\left[ B_{R}^{u}\right] _{32}=Q_{R}^{u}\left( x-1\right)
s_{t}s_{c}A\lambda ^{2}
\end{equation}
\begin{equation}
\left[ B_{L,R}^{l}\right] _{33}=Q_{L,R}^{l}.
\end{equation}
Then
\begin{equation}
B_{1}=2\left( x-1\right) ^{2}A^{2}\lambda ^{4}\left[ \left(
Q_{L}^{u}Q_{L}^{l}\right) ^{2}+\left( Q_{R}^{u}Q_{R}^{l}\right)
^{2}\right] ,
\end{equation}
\begin{equation}
B_{2}=2\left( x-1\right) ^{2}A^{2}\lambda ^{4}\left[ \left(
Q_{L}^{u}Q_{R}^{l}\right) ^{2}+\left( Q_{R}^{u}Q_{L}^{l}\right)
^{2}\right],
\end{equation}
\begin{equation}
B_{3}=Q_{R}^{u}Q_{L}^{l}\left( x-1\right) ^{3}s_{t}s_{c}A^{3}\lambda
^{6} \left[ \left( Q_{L}^{u}\right) ^{2}+\left( Q_{R}^{u}\right)
^{2}\right],
\end{equation}
\begin{equation}
B_{4}=Q_{R}^{l}Q_{L}^{u}\left( x-1\right) A\lambda ^{2}\left(
\left\vert Q_{L}^{l}\right\vert ^{2}+\left\vert Q_{R}^{l}\right\vert
^{2}\right),
\end{equation}
and
\begin{equation}
B_{5}=4s_{t}s_{c}Q_{L}^{u}Q_{R}^{u}Q_{L}^{l}Q_{R}^{l}\left(
x-1\right) ^{2}A^{2}\lambda ^{4}
\end{equation}
For scenario iii)
\begin{equation}
\left[ B_{L}^{u}\right] _{32}=Q_{L}^{u}\left(
x_{1}V_{td}V_{cd}^{\ast }+x_{2}V_{ts}V_{cs}^{\ast
}+x_{3}V_{tb}V_{cb}^{\ast }\right) ,
\end{equation}
\begin{equation}
\left[ B_{R}^{u}\right] _{32}=0,
\end{equation}
\begin{equation}
\left[ B_{L,R}^{l}\right] _{33}=Q_{L,R}^{l}.
\end{equation}
Then
\begin{eqnarray}
B_{1} &=&2\left\vert Q_{L}^{u}\right\vert ^{2}\left\vert
Q_{L}^{l}\right\vert ^{2}\left[ \left( x_{2}-x_{1}\right)
^{2}\left\vert V_{ts}\right\vert ^{2}\left\vert V_{cs}\right\vert
^{2}+\left( x_{3}-x_{1}\right) ^{2}\left\vert V_{tb}\right\vert
^{2}\left\vert V_{cb}\right\vert ^{2}\right. \nonumber\\
&&\left. +\left(
x_{2}-x_{1}\right) \left( x_{3}-x_{1}\right) V_{ts}V_{cs}^{\ast
}V_{tb}^{\ast }V_{cb}+\left( x_{2}-x_{1}\right) \left(
x_{3}-x_{1}\right) V_{ts}^{\ast }V_{cs}V_{tb}^{\ast }V_{cb}\right],
\end{eqnarray}
\begin{eqnarray}
B_{2} &=&2\left\vert Q_{L}^{u}\right\vert ^{2}\left\vert
Q_{R}^{l}\right\vert ^{2}\left[ \left( x_{2}-x_{1}\right)
^{2}\left\vert V_{ts}\right\vert ^{2}\left\vert V_{cs}\right\vert
^{2}+\left( x_{3}-x_{1}\right) ^{2}\left\vert V_{tb}\right\vert
^{2}\left\vert
V_{cb}\right\vert ^{2}\right.\nonumber\\
&&\left. +\left( x_{2}-x_{1}\right) \left( x_{3}-x_{1}\right)
V_{ts}V_{cs}^{\ast }V_{tb}^{\ast }V_{cb}+\left( x_{2}-x_{1}\right)
\left( x_{3}-x_{1}\right) V_{ts}^{\ast }V_{cs}V_{tb}^{\ast
}V_{cb}\right],
\end{eqnarray}
\begin{equation}
B_{3}=0,
\end{equation}
\begin{equation}
B_{4}=Q_{R}^{l}Q_{L}^{u}\text{Re}\left( \left( x_{2}-x_{1}\right)
V_{ts}V_{cs}^{\ast }+\left( x_{3}-x_{1}\right) V_{tb}V_{cb}^{\ast
}\right) \left( \left\vert Q_{L}^{l}\right\vert ^{2}+\left\vert
Q_{R}^{l}\right\vert ^{2}\right)
\end{equation}
and
\begin{equation}
B_{5}=0.
\end{equation}

\subsection{Integrals}
\begin{table}
\centering
\begin{tabular}{|c|c|c|c|}
  \hline
  $I_i$ & $m_e$ & $m_{\mu}$ & $m_\tau$\\
  \hline
  $I_1$ &0.4999 &0.4990 &0.4887 \\
  \hline
  $I_2$ &0.3333 &0.3332 &0.3315 \\
  \hline
  $I_3$ &0.25 &0.25 &0.2499\\
  \hline
  $I_4$ &0.3333 &0.3324 &0.3229 \\
  \hline
  $I_5$ &0.2499 &0.2491 &0.2402 \\
  \hline
\end{tabular}
  \caption{Numerical values of the integrals for each charged lepton.}
  \label{tableIs}
\end{table}
\begin{table}
\centering
\begin{tabular}{|c|c|c|c|}
  \hline
  $a_i$ & $m_e$ & $m_{\mu}$ & $m_\tau$\\
  \hline
  $a_1$&$0.0833$ &$0.0833$ & $0.0832$\\
  \hline
  $a_2$ &$0.0833$ &$0.0833$ &$0.0832$ \\
  \hline
  $a_3$ &$5.823\times10^{-12}$ &$2.489\times10^{-7}$ &$7.014\times10^{-5}$ \\
  \hline
  $a_4$ &$-2.487\times10^{-3}$ &$-2.487\times10^{-3}$ &$-2.485\times10^{-3}$ \\
  \hline
  $a_5$& $-1.303\times10^{-13}$&$-5.571\times10^{-9}$ &$-1.571\times10^{-6}$ \\
  \hline
\end{tabular}
  \caption{Numerical values of the $B_i$, $i=1,...,5$.}
  \label{tableB}
\end{table}

\begin{equation}
I_{1}=\frac{1}{2}(1-\mu _{1}^{2})+\mu _{1}\ln \mu _{1},
\end{equation}
\begin{equation}
I_{2}=\frac{1}{3}\left( 1-\mu _{1}\right) ^{3},
\end{equation}
\begin{equation}
I_{3}=\frac{1}{4}\left( 1-\mu _{1}\right) \left( 1+\mu _{1}+\mu
_{1}^{2}+\mu _{1}^{3}\right) +\mu _{1}^{2}\ln \mu _{1},
\end{equation}
\begin{equation}
I_{4}=\frac{1}{6}\left( 1-\mu _{1}\right) \left(
2+5\mu _{1}-\mu _{1}^{2}\right) +\mu _{1}\ln \mu _{1},
\end{equation}
\begin{equation}
I_{5}=\frac{1}{12}\left( 1-\mu _{1}\right) \left( \mu _{1}^{3}-5\mu
_{1}^{2}+13\mu _{1}+3\right) +\mu _{1}\ln \mu _{1}.
\end{equation}
\begin{equation}
a_{1}=\left( 1-\mu _{3}\right) I_{2}-I_{3},
\end{equation}
\begin{equation}
a_{2}=\left( 1-\mu _{3}\right) I_{4}-I_{5},
\end{equation}
\begin{equation}
a_{3}=2\mu _{1}\left( 2I_{1}-I_{2}-I_{4}\right) ,
\end{equation}
\begin{equation}
a_{4}=2\sqrt{\mu _{3}}\left[ \left( 1+2\mu _{1}-\mu _{3}\right)
I_{1}-I_{2}-I_{4}\right] ,
\end{equation}
\begin{equation}
a_{5}=-4\sqrt{\mu _{3}}\mu _{1}I_{1},
\end{equation}
\begin{equation}
B_{1}=2\left( \left\vert B_{L}^u\right\vert ^{2}\left\vert
B_{L}^l\right\vert ^{2}+\left\vert B_{R}^u\right\vert ^{2}\left\vert
B_{R}^l\right\vert ^{2}\right)
\end{equation}
\begin{equation}
B_{2}=2\left( \left\vert B_{L}^u\right\vert ^{2}\left\vert
B_{R}^l\right\vert ^{2}+\left\vert B_{L}^l\right\vert ^{2}\left\vert
B_{R}^u\right\vert ^{2}\right)
\end{equation}
\begin{equation}
B_{3}=\textrm{Re}\left( B_{L}^l B_{R}^{u\ast }\right) \left(
\left\vert B_{L}^u\right\vert ^{2}+\left\vert B_{R}^u\right\vert
^{2}\right)
\end{equation}
\begin{equation}
B_{4}=\textrm{Re}\left( B_{L}^u B_{R}^{l\ast }\right) \left(
\left\vert B_{L}^l\right\vert ^{2}+\left\vert B_{R}^l\right\vert
^{2}\right)
\end{equation}
and
\begin{equation}
B_{5}=2\left[ \textrm{Re}\left( B_{L}^u B_{L}^l B_{R}^{u\ast}
B_{R}^{l\ast }\right) + \textrm{Re}\left( B_{L}^u B_{L}^{l\ast
}B_{R}^u B_{R}^{l\ast }\right) \right] .
\end{equation}

\end{document}